\documentclass[manuscript,screen]{acmart}
 
\AtBeginDocument{%
  \providecommand\BibTeX{{%
\normalfont B\kern-0.5em{\scshape i\kern-0.25em b}\kern-0.8em\TeX}}}
 
\setcopyright{rightsretained}
\copyrightyear{2024}
\acmYear{2024}
\acmDOI{}
 
\acmConference[CCAI 2024]{CHI 2024 Workshop on Child-centred AI Design}{May 11, 2024}{Honolulu, HI, USA}
 
\acmBooktitle{CHI 2024 Workshop on Child-centred AI Design, May 11, 2024, Honolulu, HI, USA}

\usepackage{caption}
\usepackage{subcaption}
\usepackage{_macros}
\usepackage{multirow}
\usepackage{wrapfig}

\textfloatsp{2mm}

\begin{document}

\title{Designing AI-Enabled Games to Support Social-Emotional Learning \\ for Children with Autism Spectrum Disorders}

\author{\href{https://orcid.org/0009-0008-7730-8552}{Yue Lyu}}
\email{yue.lyu@uwaterloo.ca}
\affiliation{
    \institution{University of Waterloo}
    \country{Canada}
}

\author{\href{https://orcid.org/0000-0002-7705-2031}{Pengcheng An}}
\email{anpc@sustech.edu.cn}
\affiliation{
    \institution{Southern University of Science and Technology}
    \country{China}
}

\author{\href{https://orcid.org/0009-0001-0373-8241}{Huan Zhang}}
\email{h648zhang@uwaterloo.ca}
\affiliation{
    \institution{University of Waterloo}
    \country{Canada}
}

\author{\href{https://orcid.org/0000-0002-9642-9666}{Keiko Katsuragawa}}
\email{kkatsuragawa@uwaterloo.ca}
\affiliation{
    \institution{National Research Council}
    \country{Canada}
}
\affiliation{
    \institution{University of Waterloo}
    \country{Canada}
}

\author{\href{https://orcid.org/0000-0001-5008-4319}{Jian Zhao}}
\email{jianzhao@uwaterloo.ca}
\affiliation{
    \institution{University of Waterloo}
    \country{Canada}
}

\renewcommand{\shortauthors}{Lyu et al.}

\begin{abstract}
Children with autism spectrum disorder (ASD) experience challenges in grasping social-emotional cues, which can result in difficulties in recognizing emotions and understanding and responding to social interactions.
Social-emotional intervention is an effective method to improve emotional understanding and facial expression recognition among individuals with ASD.
Existing work emphasizes the importance of personalizing interventions to meet individual needs and motivate engagement for optimal outcomes in daily settings. 
We design a social-emotional game for ASD children, which generates personalized stories by leveraging current advancement of artificial intelligence. %
Via a co-design process with five domain experts, this work offers several design insights into developing future AI-enabled gamified systems for families with autistic children. 
We also propose a fine-tuned AI model and a dataset of social stories for different basic emotions.

\end{abstract}



\maketitle

\section{Introduction}
Autism spectrum disorder (ASD) is a complex neurodevelopmental disorder characterized by impairments in social-emotional reciprocity, such as challenges with recognizing and expressing emotions and difficulties in understanding and responding to social interactions \cite{DSM-5, Shah2014}.
Such challenges are affecting approximately 1 in 100 children worldwide \cite{world-health-organization}, with the number continuously growing \cite{dinstein2023two, li2022global}. %
Social-emotional interventions encompass various therapeutic approaches, aiming to improve social-emotional skills, such as emotional understanding or processing among individuals with ASD.
They have shown promise in aiding children with ASD to better comprehend emotions \cite{Golan2010}, enhance facial emotion recognition (FER) skills \cite{Tanaka2010}, and improve their response to others' emotions \cite{Bauminger2002}, thereby fostering more meaningful social interactions and relationships \cite{Reichow2010}.

However, caregivers' active participation in interventions for children with ASD is hindered by financial constraints and limited access to services and resources \cite{rogge_economic_2019, rios_exploring_2020}; and few studies have looked into developing intervention tools to promote convenience and affordability to families with autistic children, to encourage participation, and ultimately contribute to improved children's social-emotional development \cite{Reichow2012, Karst2012}. 
There is a need for the design and development of tools to facilitate caregivers in children's social-emotional development \cite{Gengoux2019}, which may leverage existing evidence-based interventions.
Social stories are an effective method for social-emotional interventions, offering a structured way to teach social-emotional skills \cite{Qi2018}, including emotional understanding and recognition \cite{Bader2006} to autistic children.
For instance, a social story might depict a scenario where a friend is crying due to sadness, and subsequently outline suitable reactions, aiding autistic children in understanding and reacting to various emotions.
Despite their effectiveness, it is a challenge to create social stories efficiently and personalized to fit the diverse personal needs and preferences of autistic children \cite{Kokina2010}. 

Artificial Intelligence (AI) has shown substantial promise in enhancing diagnostics  \cite{sabzevari2023artificial}, face recognition and monitoring \cite{bhargavi2023ai}, and educational outcomes \cite{chen2021cognitive} for individuals with ASD.
For instance, AI enables fully automated interactive question answering between children and a chatbot \cite{zhang2022storybuddy}.  
AI may also generate language that is easy for autistic children to understand, ensuring the clarity and effectiveness of the stories \cite{koegel_natural_1987, krantz_teaching_nodate}. 
While AI has the potential to assist in emotional facial expressions recognition and content generation, saving valuable time and resources, little HCI research has been done to explore it in the context of ASD intervention.

To address this gap, we carried out a design of an AI-enabled mobile game that could provide autistic children with personalized stories in social-emotional interventions. %
As a necessary step of verifying our design ideas, we conducted a co-design process with five domain experts in ASD and neuroscience. 
In this work-in-progress paper, we report the obtained design insights into the development of future AI-enabled gamified systems for families with autistic children.  
We also propose a fine-tuned model based on GPT-4 \cite{OpenAI2024} and a dataset of diverse social stories with various emotions, which set the foundation for developing social-emotional games for ASD children.

\section{Co-design Process}
We employed a co-design approach to verify our game design ideas by working with domain experts (E1-E5).
E1 is a university researcher specializing in emotion-related neuroscience. 
E2 and E3 are teachers from the special education center for autistic children. 
E4 and E5 are experts in education for autistic children.
We closely communicated and collaborated with the experts via various forms, including questionnaires, interviews, co-design sessions, and remote online discussions. 
During this process, we acquired knowledge about ASD and autistic children, understood current practices and challenges of social-emotional learning, and identified key principles for optimizing the use of AI for effective interventions.

First, we carried out two semi-structured interviews with two of the experts (E4 and E5), which led us to choose an easy-setup mobile platform to access AI and design insights of design children-centred games to promote engagement and intervention outcomes.
Next, we proposed a social stories dataset generated by a fine-tuned model using GPT-4 \cite{OpenAI2024}, which was carefully crafted through an examination of literature \cite{reynhout_social_2006, gray_social_1993}, coupled with insights from three of the experts (E1-E3).
Last, we used the model to generate a story dataset covering seven basic emotions (\ie, happiness, surprise, sadness, fear, disgust, anger, and neutral), and asked experts to rate each story and provide feedback.

\subsection{Design Insights}  \label{expert-insights}
Based on our co-design process with the experts, we derived the following insights that could shed light on the future development of social-emotional games for autistic children. 

\begin{enumerate}[leftmargin=8mm, label=\textbf{I\arabic*}]
\item[\textbf{I1}] \textbf{Adaptive Story Levels:} Generate stories by emotion and scenario complexity to meet the children's ability level for enhanced social-emotional learning, requiring richer backgrounds for higher levels.
\item[\textbf{I2}] \textbf{Familiar Story Scenarios:} Use situations that children are familiar with for specific emotion experiences that are close the real world, aiding the understanding of complex emotions through experience.
\item[\textbf{I3}] \textbf{Diverse Story Generation:} Generate various story contents and characters, offering multiple aspects for comprehensive emotion understanding.
\item[\textbf{I4}] \textbf{Attractive Visual Illustrations:} Emphasize necessary visual stimuli in stories, crucial for engaging children and supporting emotional learning.
\item[\textbf{I5}] \textbf{Multimodal Gamified Feedback:} Prioritize visual and haptic channels for autistic children's engagement, integrating visual cues dynamically and smoothly to avoid distraction.
\end{enumerate}

\subsection{Emotional Story Generation}
Following the writing guideline by Gray \cite{gray1993social}, our approach integrates a variety of sentence types—descriptive, perspective, directive, control, affirmative, and cooperative—each chosen for its unique contribution to the narrative's effectiveness. 
To enrich our stories with professional insights, we combined these structured sentence types with experts' insights in prompts. 
Utilizing AI technologies, specifically GPT-3.5 and GPT-4, we created stories of the seven basic emotions and gathered further suggestions from experts for crafting the stories tailored to ASD children.
Incorporating the following received suggestions into our prompts, we refined our model to deliver content that more closely aligns with the needs of autistic children, aiming for the best possible outcomes.
\begin{itemize} 
 \item \textbf{Life-like Environments:} Including items and settings that closely resemble real life in the stories can enhance students' life experiences, which highlights the importance of incorporating realistic elements into stories.
 \item \textbf{Accessibility for Different Abilities:} Stories should be adaptable for students with varying degrees of autism, including those without speech functions, who may require more guidance through language and visual cues.
 \item \textbf{Simplicity in Emotional Triggers:} The causes of emotions in stories should be straightforward and instinctual, such as disgust from dirtiness or happiness from recovering a lost item, reflecting basic human nature.
 \item \textbf{Language Use:} The vocabulary should be simple, with concrete nouns for lower levels and possibly abstract terms for higher levels, but always kept understandable. For students with severe autism who lack verbal skills, alternative communication methods like pointing can be used to express answers or preferences.
 \item \textbf{Consistency in Emotion and Preferences:} In stories of lower difficulty, maintaining consistent attitudes and preferences is advised to avoid confusing students with different emotional responses.
 \item \textbf{Sentence Structure:} Longer sentences should be broken down to aid students in thinking and expressing themselves more clearly.
\end{itemize}

\subsection{Emotional Story Dataset}

\begin{table*}[h]
\small
\centering
\caption{Experts' ratings on generated emotional stories on a scale from 1 - very inappropriate to 7 - very appropriate}
\vspace{-3mm}
\begin{tabular}{lrrrr}
 \toprule
 \textbf{Emotion} & \textbf{E1} & \textbf{E2}  & \textbf{E3} & \textbf{Mean} \\ 
 \midrule
Anger & 6.88  & 5.13 & 5.38 & 5.79\\ 
Happiness & 6.00 & 5.75 & 4.63 & 5.46 \\
Sadness& 4.75  & 5.50 & 4.63 &  4.96 \\ 
Disgust&  6.50 &4.38 & 5.00 & 5.29\\ 
Fear &  3.63 & 4.88 & 4.50& 4.33\\  
Surprise & 5.25 & 5.75 & 4.63 & 5.21  \\ 
Neutral & 5.50  & 4.88 & 5.12 & 5.17  \\ 
\midrule
Mean& 5.50 &	5.18&	4.84&	5.17\\
\bottomrule
\label{tab:stories-rates}
\end{tabular}
\end{table*}

Leveraging the enhanced model, we produced a social story dataset encompassing all the basic emotions, embedding scenarios (\eg, birthdays, school, and home) and themes (\eg, pets and drawings) rooted in real-life scenarios. 
We used the model to generate 56 stories, 8 for each emotion, which were then assessed by experts using a 7-point scale %
to determine their suitability, and the results are shown in \autoref{tab:stories-rates}.
The overall mean rating across all emotions and experts was approximately 5.17, indicating a generally favourable assessment of the stories' suitability for the social-emotional intervention.

\section{Envision of Social-Emotional Game Development}
Through the co-design process with ASD experts, the insights that we distilled as well as the story generation model and dataset that we produced have set a solid foundation for our future development of the AI-enabled mobile game providing personalized stories in social-emotional interventions. 
We envision that the game encompasses three distinct phases, starting with an AI-enabled customization phase that generates tailored social stories for ASD children, followed by another phase for emotional understanding with caregivers. 
Then, interactive turn-taking between caregivers and children, enhanced by AI-driven facial expression recognition, is presented to allow children to observe and mimic facial expressions effectively.
After the development of the game, we plan to carry out an evaluation with ASD children and their caregivers and future derive insightful empirical knowledge to inform future research on similar topics.

\bibliographystyle{ACM-Reference-Format}
\bibliography{bib/references.bib,bib/zotero}

\end{document}